# The impact of interfacial chemistry on the band offset of GaAs/$Ga_2O_3$ heterostructures

Sofia Apergi,[1,*] Alfredo Pasquarello,[2] Charles Cornet,[1] and Laurent Pedesseau[1,†]

[1]Univ Rennes, INSA Rennes, CNRS, Institut FOTON – UMR 6082, F-35000 Rennes, France
[2]Chaire de Simulation à l'Echelle Atomique (CSEA), Ecole Polytechnique Fédérale de Lausanne (EPFL), CH-1015 Lausanne, Switzerland

**ABSTRACT**. $Ga_2O_3$/GaAs heterojunctions are emerging as promising candidates for next-generation power electronics, photonics, and energy devices, leveraging the high breakdown voltage and thermal stability of $Ga_2O_3$ alongside the mature technology, high hole mobility, and higher refractive index of GaAs. The efficiency of these devices depends strongly on the band alignment between the two materials, however both type-I and type-II alignment have been reported in the literature for these heterostructures. To address this ambiguity, we use hybrid density functional theory to systematically investigate the band alignment at GaAs/$Ga_2O_3$ interfaces, focusing on the role of interface chemistry. By considering Ga-O-, As-, and As-O-rich interfaces both in amorphous and crystalline $Ga_2O_3$ phases, we demonstrate that interface stoichiometry determines the alignment type: Ga-O-rich interfaces exhibit type-II alignment with large valence band offsets (~3.1 eV), while As-rich and As-O-rich interfaces favor type-I alignment with reduced offsets (~2.3–2.6 eV). These trends are attributed to interface dipole formation driven by bonding configuration. Our findings provide insight into the relationship between chemistry and band alignment in GaAs/$Ga_2O_3$ heterostructures, enabling targeted optimization for specific device applications.

Heterojunctions between gallium oxide ($Ga_2O_3$) and gallium arsenide (GaAs) are attracting increasing attention for their potential application in power electronics, photonics, and energy devices. Especially, due to its high breakdown voltage, thermal stability, and cost-effective fabrication, the wide band gap semiconductor $Ga_2O_3$ is explored extensively for replacing established semiconductors, such as GaN and SiC, in high power electronics [1]. However, the persistent challenge of achieving p-doping in $Ga_2O_3$ has limited the use of this material to mainly unipolar devices, such as Schottky barrier diodes (SBDs) [2,3] and metal-oxide-semiconductor field-effect transistors (MOSFETs) [4]. The integration of $Ga_2O_3$ with GaAs offers a promising pathway for overcoming this limitation, by taking advantage of the superior properties of the latter, such as its high hole mobility and overall technological maturity [5]. Additionally, the ability to easily grow $Ga_2O_3$ on III-V substrates like GaAs via a variety of fabrication methods such as molecular beam epitaxy [6] or grafting [7–9], provides a significant degree of control over the formed interface. As a consequence, GaAs/$Ga_2O_3$ heterostructures are being explored for more applications, such as dual-band photodetectors [10], light-emitting diodes [11,12], and gas sensing [13].

Interfaces between $Ga_2O_3$ and GaAs are also inherently present in many GaAs-based devices. As with other III-V semiconductors, GaAs surfaces readily form a native oxide under a variety of environmental and processing conditions. While the exact composition and degree of crystallinity of this oxide strongly depend on the fabrication conditions and the environment, $Ga_2O_3$ is generally recognized as its dominant constituent. Considerable quantities of $As_2O_3$ may also form initially, however, being volatile and water-soluble, this species is eventually removed from the native oxide layer, either spontaneously or with appropriate treatment such as rinsing with water [14–17]. As $Ga_2O_3$ can serve as a passivation layer for GaAs, the intentional oxidation of the semiconductor surface has been explored, as a means of better controlling the material properties [18].

One of the most critical properties of heterojunctions is the band alignment between the constituent layers. The type of alignment and the magnitude of the valence and conduction band offsets (VBO and CBO, respectively) dictate both the direction of charge transport and the efficiency of charge separation at the interface. While the valence band maximum (VBM) of $Ga_2O_3$ lies below that of GaAs, there have been conflicting reports regarding the CBO of the two materials. Specifically, as their conduction band minima (CBM) are energetically similar, both type-I and type-II heterostructures have been realized, suggesting that the alignment type of GaAs/$Ga_2O_3$ heterostructures is a tunable parameter [5,7–10,19–21]. However, designing high-performance devices requires a thorough understanding of the factors governing this band alignment.

*Contact author: sofia.apergi@insa-rennes.fr
†Contact author: laurent.pedesseau@insa-rennes.fr

In this work, we use advanced density functional theory (DFT) methodologies to investigate the band alignment at $GaAs/Ga_2O_3$ heterostructures, focusing on the impact of interface chemistry, which has been shown to strongly influence band alignment [22,23]. To circumvent the challenges associated with the lattice mismatch between GaAs and crystalline $Ga_2O_3$, we model heterostructures using an amorphous oxide layer. The band alignment between GaAs and crystalline $Ga_2O_3$ is then calculated by aligning the electronic levels of the different oxide phases, amorphous and crystalline, from bulk calculations. We study three model interfaces, namely a Ga-O-rich, an As-rich, and an As-O-rich interface, and compare their band lineups. Our findings confirm that interface chemistry significantly impacts the band alignment, with the Ga-O-rich interface giving rise to type-II alignment, while the As-rich and As-O-rich interfaces most likely lead to type-I alignment, with VBOs varying within a ~0.8 eV range for different interface compositions.

Density functional theory (DFT) calculations were performed using the freely available CP2K/Quickstep package [24]. Analytic Goedecker-Teter-Hutter (GTH) pseudopotentials were used to represent the core electrons [25]. A triple-ζ correlation-consistent polarized basis set (cc-pVTZ) was used for O and As, and a double-ζ (cc-pVDZ) for Ga [26]. The plane wave cutoff for the electron density was set to 600 Ry. The Perdew, Burke, and Ernzerhof (PBE) exchange-correlation functional within the generalized gradient approximation (GGA) [27] was used for the structural relaxation of the simulation cells. The Brillouin zone was sampled using a 6×6x6 and a 7×7×5 **k**-point grid for GaAs and $\beta$-$Ga_2O_3$ respectively, yielding the lattice parameters presented in Table I, in good agreement with experimental references. For the amorphous $Ga_2O_3$ (am-$Ga_2O_3$), we start from a unit cell corresponding to amorphous $Al_2O_3$ containing 160 atoms from Ref. [28], and we replace Al with Ga before fully relaxing using a 3×3×3 **k**-point grid. We then constrain both the $x$ and $y$ dimensions to 11.5 Å to match those of the GaAs 2×2 substrate and relax again, this time only along the $z$ axis, while also optimizing the atomic positions. This leads to an amorphous oxide with a density of 5.61 $g/cm^3$. Experimentally, the density of am-$Ga_2O_3$ varies significantly depending on fabrication conditions, particularly temperature, and can range from 3.9 to 5.6 $g/cm^3$ [29].

For a more accurate description of the electronic properties, we use the hybrid PBE0 functional [30], and we set the parameter $\alpha$ that controls the fraction of Fock exchange to 0.27, as this value yields predictions for the band gaps of the crystalline semiconductors close to the experimental values. Specifically, we calculate a band gap of 1.53 eV for GaAs, compared to the experimental value of 1.42 eV, and 4.77 eV for $\beta$-$Ga_2O_3$, well within the experimentally reported range. Using this approach, we derive a band gap of 4.18 eV for am-$Ga_2O_3$, which also falls within the corresponding experimental range of 4.1-4.4 eV and aligns well with the recent theoretical work of Kaewmeechai *et al*. [29]. In this work, the narrowing of the band gap in the amorphous, compared to the crystalline phase, was attributed to localized states near the band edges, originating from the structural disorder. It should be noted that the values in Table I were calculated at 0 K, while the experimental ranges typically correspond to room temperature. For the materials studied, the band gap is expected to narrow by approximately 0.1-0.2 eV from 300 K to 0 K, while the change in density is expected to be negligible.

In Fig. 1, the projected density of states (PDOS) for the two studied $Ga_2O_3$ phases is presented. Both phases exhibit similar features with mainly O contributions for the VBM and a CBM of mixed Ga-O character. The same observations were made by Kaewmeechai *et al.*, using models built with the PBE0-TC-LRC functional [29]. We align the electronic levels of the two phases using the average

TABLE I. Calculated lattice parameters, mass density $\rho$, and band gap $E_g$ of GaAs, $\beta$-$Ga_2O_3$, and am-$Ga_2O_3$ along with experimental values, when applicable.

| | $a$ (Å) | $b$ (Å) | $c$ (Å) | $\rho$ ($g/cm^3$) | $E_g$ (eV) |
|---|---|---|---|---|---|
| | | | **GaAs** | | |
| This work | 5.74 | | | | 1.53 |
| Expt. | 5.65 [31] | | | | 1.42 [32] |
| | | | **$\beta$-$Ga_2O_3$** | | |
| This work | 12.36 | 3.05 | 5.64 | 5.86 | 4.77 |
| Expt. | 12.23 [33] | 3.04 [33] | 5.8 [33] | 5.77 | 4.26-5.24 [34] |
| | | | **am-$Ga_2O_3$** | | |
| This work | 11.5 | 11.5 | 13.43 | 5.61 | 4.18 |
| Expt. | | | | 3.9-5.6 [29] | 4.1-4.4 [29] |

*Contact author: sofia.apergi@insa-rennes.fr
†Contact author: laurent.pedesseau@insa-rennes.fr

2*s* level of O atoms. We thus find that the narrowing of the band gap in the amorphous phase compared to the crystalline one, comes from a broadening of both valence and conduction bands by approximately 0.3 eV.

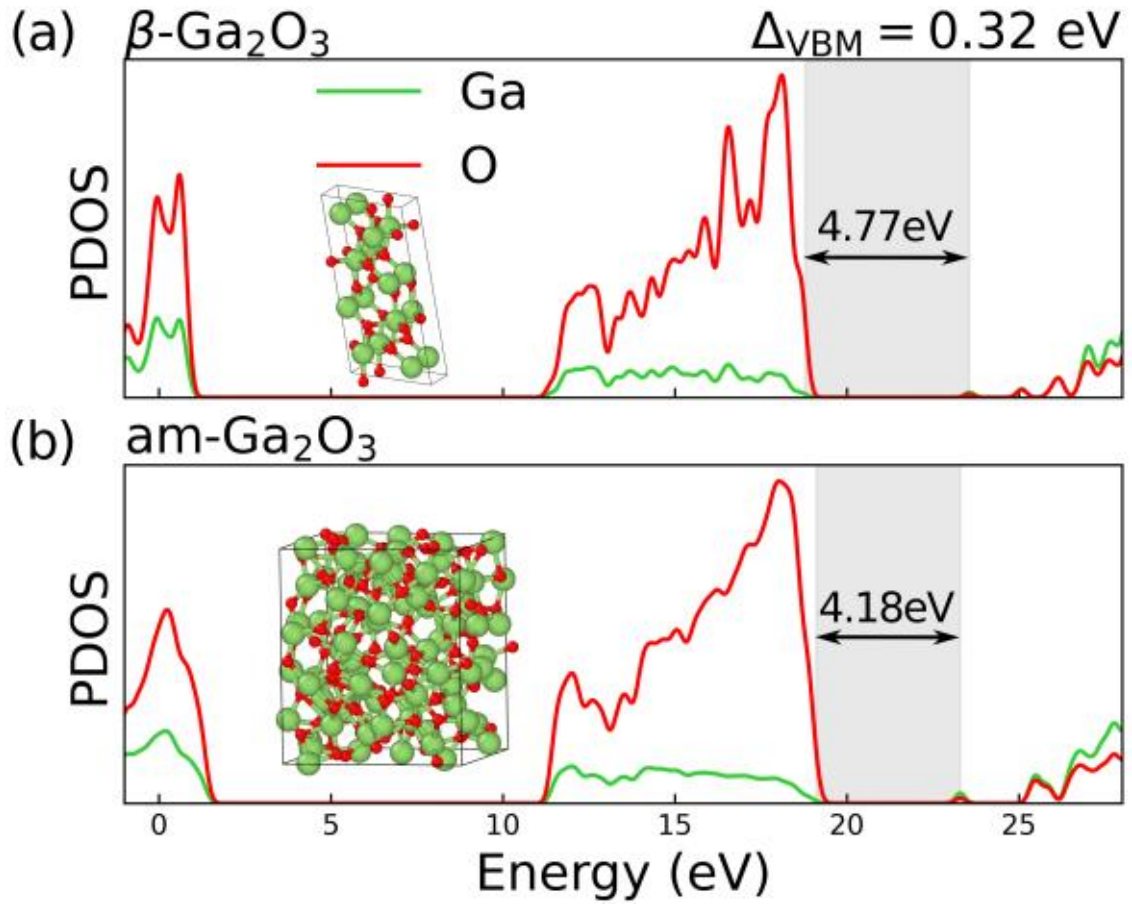

FIG 1. Element projected electronic density of states (PDOS) for (a) $\beta$-$Ga_2O_3$ and (b) am-$Ga_2O_3$ obtained from calculations using the PBE0 functional with $\alpha$ = 0.27. The energies are referred to the average 2*s* level of O atoms. $\Delta_{VBM}$ indicates the valence band offset between $\beta$-$Ga_2O_3$ and am-$Ga_2O_3$.

Next, we proceed with determining the band alignment between GaAs and $Ga_2O_3$. For this, we consider three different model structures of the interface namely a Ga-O-rich, an As-rich, and an As-O-rich model. A vacuum region of ~60 Å is included in all heterostructures to separate the periodic images along the direction perpendicular to the interface. These model structures are relaxed using the PBE functional, while the Brillouin zone is sampled only at the Γ-point, leading to the interface structures shown in Fig. 2. These models consist of 17-layer GaAs slabs (16 layers for the Ga-rich case), approximately 24 Å thick and oriented along the (001) direction. To passivate the bottom surface, which is unstable due to the polarity associated with this orientation, we employ the As-rich c(4×4)$\beta$ reconstruction, which has been determined as one of the most stable ones [35]. To construct the Ga-rich model, the 16-layer Ga-terminated GaAs slab is interfaced with a ~12 Å thick am-$Ga_2O_3$ slab, which is created by cleaving the 160-atom cell described above. After relaxing this heterostructure, Ga-Ga dimers form on the GaAs side of the interface region. While a Ga-rich interface could in principle be studied, the existence of such dimers is not very likely, as Ga atoms readily bond to As atoms or to the much more electronegative O atoms, when these atoms are available in the environment [36]. We therefore add excess O atoms around the Ga-Ga dimers and relax the structure again, leading to the interface geometry shown in Fig. 2(a). We construct the As-rich model in a similar manner but starting from a 17-layer As-terminated GaAs slab. After relaxing this heterostructure, a few As-Ga and As-O bonds form between GaAs and the oxide, while, similar to the Ga case, several As atoms at the interface form As-As dimers, as can be seen in Fig. 2(b). Since such dimers are known to form both at the surface of GaAs, as for instance in the reconstruction employed here for the interface between GaAs and the vacuum, as well as at interfaces between GaAs and other oxides [37], we retain this model for further analysis. Lastly, to model the As-O-rich case, we start from the As-rich interface and again add excess O atoms to saturate the As-dimers. Upon relaxation, most As-dimers broke apart, as O atoms bonded to surface As atoms and even to some subsurface Ga atoms of GaAs [Fig. 2(c)]. Indeed, O is more likely to form bonds with Ga rather than with As [36], but as the occurrence of As-O bonds cannot be ruled out, our model can serve as an extreme case [20].

In Fig. 3, we present the *z*-resolved density of states (DOS), calculated using the PBE0 functional, for the three studied heterostructures. As expected, the VBM of am-$Ga_2O_3$ lies significantly below that of GaAs in all cases, with VBOs around 2-3 eV. Notably, in the Ga-O-rich heterostructure, the CBM of the oxide is also lower than that of GaAs, resulting in a type-II band alignment. In contrast, for both the As-rich and the As-O-rich heterostructures, the CBM of am-$Ga_2O_3$ is higher than that of GaAs, yielding a type-I alignment. In addition, the CBO is slightly larger in the As-O-rich case, compared to the As-rich case.

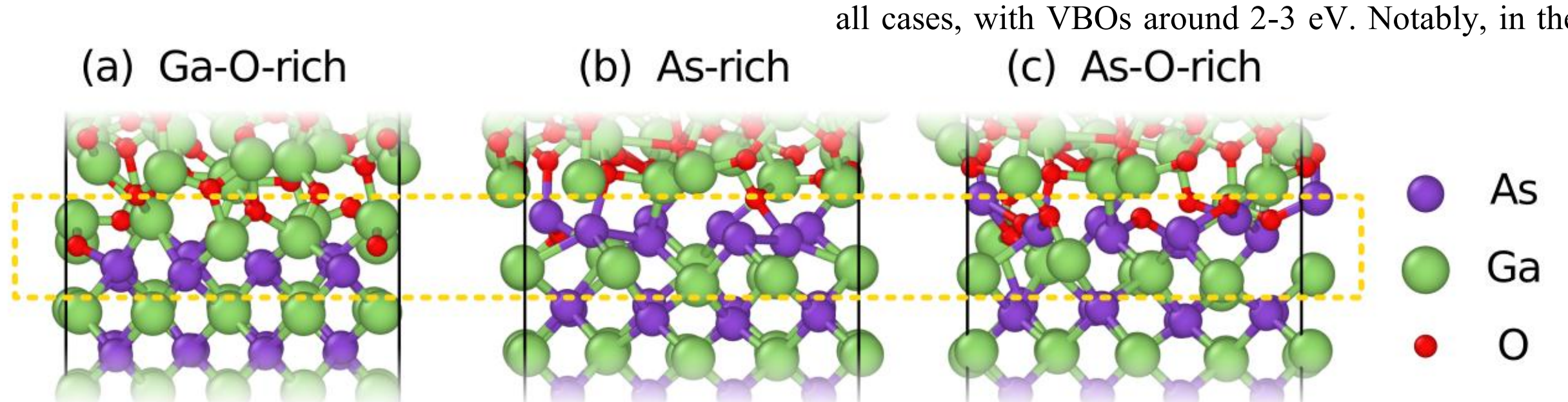

FIG 2. Atomistic representation of the relaxed interface region of (a) the Ga-O-rich, (b) As-rich, and (c) the As-O-rich models.

*Contact author: sofia.apergi@insa-rennes.fr
†Contact author: laurent.pedesseau@insa-rennes.fr

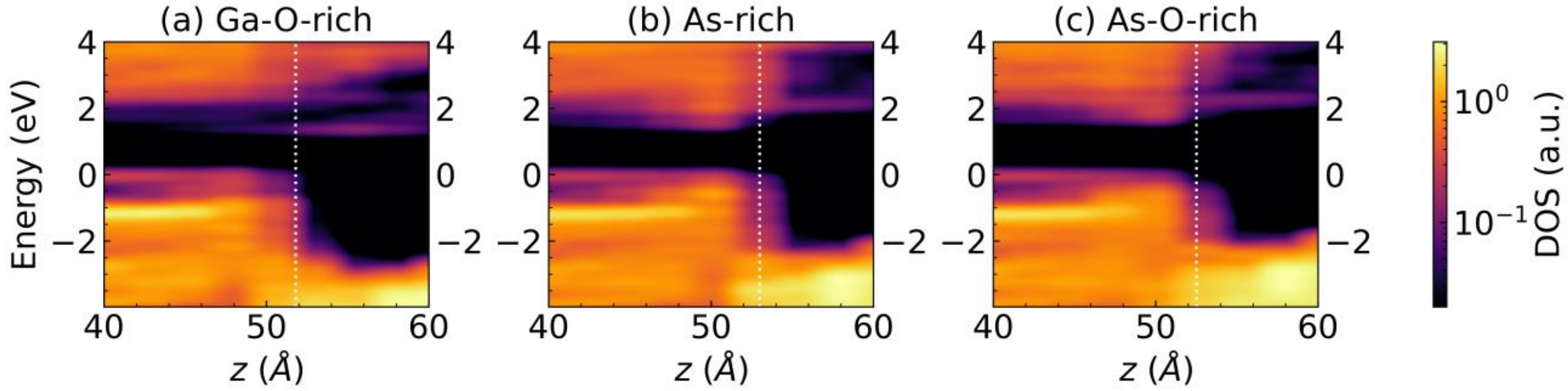

FIG 3. $z$-resolved density of states for (a) the Ga-O-rich, (b) the As-rich, and (c) the As-O-rich interface models. The white dashed line indicates the interface, while GaAs and am-$Ga_2O_3$ are to the left and right of this line, respectively. The Fermi energy has been set to zero.

For the accurate determination of the band offsets, we use the approach developed in Refs. [38] and [39]. First, we determine the VBM of bulk GaAs ($E_{\mathrm{VBM}}^{\mathrm{GaAs}}$) with respect to its average electrostatic potential ($\bar{V}^{\mathrm{GaAs}}$) and the VBM of bulk am-$Ga_2O_3$ ($E_{\mathrm{VBM}}^{\mathrm{Ga_2O_3}}$) with respect to the average level of 2$s$ O atoms ($E^{2s\,\mathrm{O}}$) of the same structure. For this, we consider bulk am-$Ga_2O_3$, relaxed as described above, with $x$ and $y$ dimensions constrained to match those of the GaAs substrate. We then identify these reference levels in the bulk-like regions of both GaAs and the oxide in the interface models, which allow us to derive the VBOs using the following expression:

$$\mathrm{VBO} = \left(E_{\mathrm{VBM}}^{\mathrm{GaAs}} - \bar{V}^{\mathrm{GaAs}}\right)_{\mathrm{bulk}} - \left(E_{\mathrm{VBM}}^{\mathrm{Ga_2O_3}} - E^{2s\,\mathrm{O}}\right)_{\mathrm{bulk}} + \left(\bar{V}^{\mathrm{GaAs}} - E^{2s\,\mathrm{O}}\right)_{\mathrm{heterostructure}} \quad (1)$$

The CBOs are then calculated as:

$$\mathrm{CBO} = \left(E_{\mathrm{g}}^{\mathrm{GaAs}} - E_{\mathrm{g}}^{\mathrm{Ga_2O_3}}\right) + \mathrm{VBO} \quad (2)$$

While the electrostatic potential could serve as a reference for the oxide as well, its amorphous nature makes the 2$s$ O levels a more reliable benchmark [40]. Specifically, within the bulk amorphous oxide, the electrostatic potential reference level can vary by ±0.25 eV from the average value, while this variation is only ±0.04 eV for the O 2$s$ reference.

Following this procedure, we obtain the band offsets presented in Table II. Consistent with the DOS of Fig. 3, the Ga-O-rich interface exhibits a type-II alignment with a VBO of 3.11 eV and a CBO of 0.46 eV. On the other hand, the As-rich interface leads to a type-I alignment, with a VBO of 2.58 eV and a CBO of –0.07 eV, with the negative sign indicating that the CBM of $Ga_2O_3$ lies closer to the vacuum level than that of GaAs in this case. Similarly, the As-O-rich interface results in a VBO of 2.37 eV and a CBO of –0.28 eV. It should be noted that as the CBO in the case of the As-rich interface is almost vanishing, its sign is sensitive to minor changes in the interface chemistry. Such small offsets have also been reported experimentally [9]. This does not affect the predicted trends in band offset variation with respect to interface chemistry, which hold regardless.

Next, we discuss the band alignment for crystalline $\beta$-$Ga_2O_3$. Due to the lattice mismatch between the two materials, creating heterostructures between GaAs and $\beta$-$Ga_2O_3$ would require either very large supercells, prohibitive for calculations with hybrid functionals, or significantly straining $\beta$-$Ga_2O_3$ to match the lattice of the GaAs substrate. However, due to the existence of many metastable $Ga_2O_3$ polymorphs, strain in $Ga_2O_3$ often leads to phase transitions [41–45], with strain thresholds for such transitions being quite low in certain cases [46,47]. Therefore, a severely strained $\beta$-$Ga_2O_3$ might not be realistic. To overcome this issue, we calculate the band offset between GaAs and crystalline $\beta$-$Ga_2O_3$ using the band offsets of the bulk oxide phases given in Fig. 1, in combination with the band lineups of the GaAs/am-$Ga_2O_3$ heterostructures. The results correspond to heterostructures between GaAs and non-strained $\beta$-$Ga_2O_3$, such as those that can result from semiconductor grafting [7]. The impact of strain, which is anticipated in epitaxially grown layers, is addressed later. The validity of this approach stems from the fact that orientation has been found to have minimal impact on the band alignment of various semiconductor interfaces, including III-V, II-VI, and oxides, compared to other properties such as interface chemistry [39,48,49]. As for the latter, while the exact bonding configuration might differ for interfaces involving crystalline rather than amorphous oxide layers, the overall trends are expected to remain meaningful.

*Contact author: sofia.apergi@insa-rennes.fr
†Contact author: laurent.pedesseau@insa-rennes.fr

As summarized in Table II, the alignment trends observed in the case of the amorphous persist, i.e. the Ga-O-rich interface exhibits a type-II alignment, while the As-rich and the As-O-rich interfaces result in type-I alignment. However, the band offsets for the crystalline phase are generally larger, except for the CBO of the Ga-O-rich case, where the CBM of $\beta$-$Ga_2O_3$ lies closer in energy to that of GaAs, compared to the CBM of am-$Ga_2O_3$. The band alignment between GaAs and the two phases of $Ga_2O_3$ for the three studied models is schematically displayed in Fig. 4.

TABLE II. Calculated VBO and CBO values for the GaAs/$Ga_2O_3$ heterostructures for both am-$Ga_2O_3$ and $\beta$-$Ga_2O_3$ using the hybrid PBE0 functional with a fraction of Fock exchange $\alpha = 0.27$.

| | am-$Ga_2O_3$ | | $\beta$-$Ga_2O_3$ | |
|---|---|---|---|---|
| | VBO | CBO | VBO | CBO |
| Ga-O-rich | 3.11 | 0.46 | 3.43 | 0.19 |
| As-rich | 2.58 | -0.07 | 2.90 | -0.34 |
| As-O-rich | 2.37 | -0.28 | 2.69 | -0.55 |

To address effect of strain on the band alignment between epitaxially grown $\beta$-$Ga_2O_3$ and GaAs, we calculate the electronic properties of the former for various degrees of biaxial strain. Specifically, we apply ±1-3% of strain along the (010) and (001) directions and optimize the bulk oxide structure, while allowing for the relaxation of the atomic positions and the cell parameter along the (100) direction. These values were chosen based on the work of Chen *et al.* [50], where metal organic chemical vapor deposition was used to grow $\beta$-$Ga_2O_3$ on (001) GaAs. The preferential growth orientation for the oxide was (400) and the lattice mismatch was determined to be ~1% and ~3% along the [010] and [001] directions respectively. Next, we calculate the shift of the band edges due to strain, compared to the unstrained oxide, using the same alignment approach as in Figure 1. The results are presented in Table III.

We can see that compressive strain generally leads to an increase of the band gap and a shift of the band edges to higher energies, while the opposite is true for tensile stress. In any case the observed trends in the band alignment still hold, even when strain is applied. However, when 1% and 3% compressive stress is applied along the (010) and (001) directions, respectively, we obtain a large CBM shift to higher energies implying that the Ga-O-rich interface would exhibit type I alignment in this case.

TABLE III. Band gap $E_g$ and position of the band edges Δ(VBM) and Δ(CBM) of biaxially strained $\beta$-$Ga_2O_3$ with respect to those of the non-strained oxide and their calculated density. The structures were relaxed by constraining the (010) and (001) directions, while allowing the relaxation of the atomic positions and the cell parameter along the (100) direction.

| | $E_g$ (eV) | Δ(VBM) (eV) | Δ (CBM) (eV) | $\rho$ (g/cm$^3$) |
|---|---|---|---|---|
| No strain | 4.77 | – | – | 5.86 |
| +1%×+1% | 4.68 | −0.06 | −0.14 | 5.80 |
| −1%×−1% | 4.84 | +0.06 | +0.14 | 5.91 |
| +1%×−1% | 4.71 | +0.06 | +0.01 | 5.86 |
| −1%×+1% | 4.79 | −0.05 | −0.03 | 5.85 |
| +1%×+3% | 4.62 | −0.14 | −0.29 | 5.75 |
| −1%×−3% | 4.86 | +0.17 | +0.27 | 5.98 |
| +1%×−3% | 4.71 | +0.18 | +0.12 | 5.92 |
| −1%×+3% | 4.69 | −0.11 | −0.18 | 5.80 |

As previously discussed, the significant impact of interface chemistry on the band alignment between two semiconductors is widely attributed to the bonding configuration at the interface and to the associated electric dipoles resulting therefrom. This effect has been demonstrated through simple models for various materials, including III-V semiconductors and oxides, by Tung and Kronik [48,49,51]. However, quantitatively estimating the amount of charge transfer across the interface directly from the macroscopically averaged charge density profiles of the heterostructures is prevented by fluctuations arising from the finite-size of our interface models. Instead, we focus directly on the interfacial atomistic structure.

Let us consider a simple model starting from the Ga-O-rich interface. We assume the interface comprises the first oxygen containing layer on the oxide side and the adjacent GaAs layer, as indicated by the yellow box in Fig. 2. In this configuration, the interface is dominated by Ga-As bonds, with the Ga atoms positioned on the oxide side. On the other hand, the As-rich interface is primarily composed of Ga-As bonds of the opposite orientation compared to the Ga-O-rich case, i.e. with As atoms on the oxide side. The associated dipole induces an upward shift of the oxide band edges and reduces the VBO compared to the Ga-O-rich interface. As for the As-O-rich case, the interface structure is similar to the As-rich case, but with some Ga-As bonds being replaced by Ga-O bonds. The higher electronegativity of O relative to As strengthens the interfacial dipoles, yielding an even smaller VBO. Bonds such as Ga-O in the Ga-O-rich interface and the As-As dimers in the As-rich interface are oriented parallel to the surface and are not expected to significantly affect the band alignment.

*Contact author: sofia.apergi@insa-rennes.fr
†Contact author: laurent.pedesseau@insa-rennes.fr

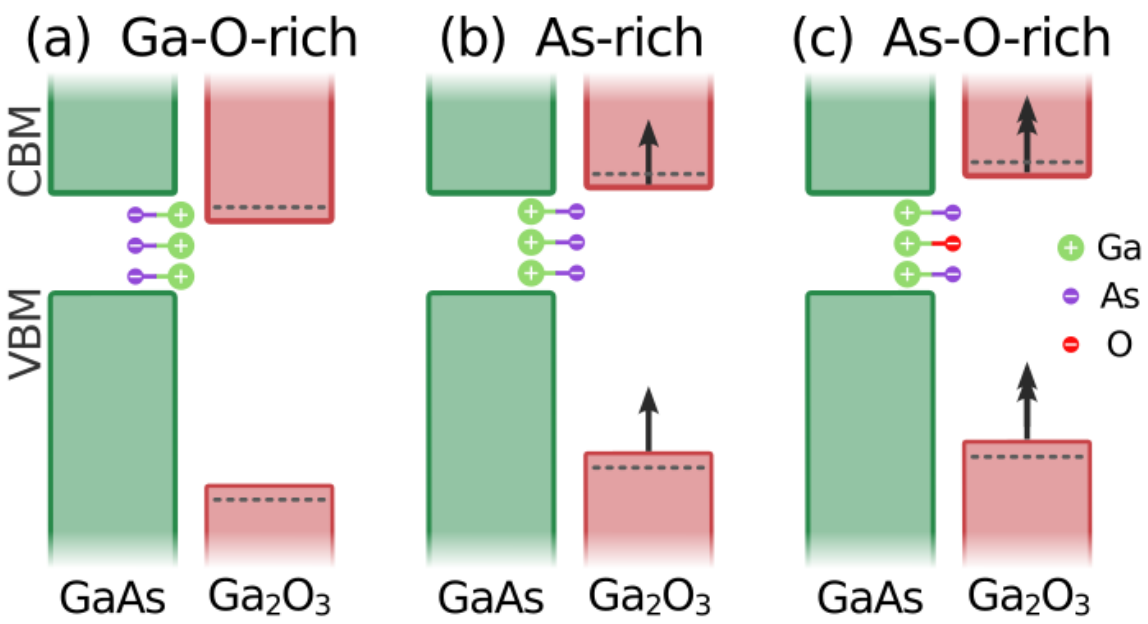

FIG 4. Schematic representation of the GaAs/$Ga_2O_3$ heterostructure band alignment for the (a) Ga-O-rich, (b) As-rich, and (c) As-O-rich interfaces. The red boxes indicate the band edges of am-$Ga_2O_3$, while the dashed gray lines indicate those of $\beta$-$Ga_2O_3$. Dipole orientation due to the bonds formed at the interface are also indicated.

GaAs/$Ga_2O_3$ heterostructures can be fabricated using a variety of methods. Most common is the growth of $Ga_2O_3$ directly on GaAs substrates by atomic layer deposition [20], molecular beam epitaxy [6], metal-organic chemical vapor deposition [41,50], or controlled thermal oxidation [52], among others. Alternatively, semiconductor grafting can be employed, where GaAs and $Ga_2O_3$ films are grown independently on separate substrates, before one is transferred to the other, and appropriate processing is performed to ensure effective bonding between the two layers [7,8]. These fabrication techniques yield oxide layers characterized by diverse degrees of crystallinity and distinct interface chemistries. The resulting heterostructure properties are further influenced by a range of fabrication parameters, such as temperature, pressure, and precursor selection, which control the chemical potential and, ultimately, the interface stoichiometry. The surface termination and orientation of the substrate are also deciding factors [53]. Generally, type-I alignment is observed in GaAs/am-$Ga_2O_3$ heterostructures [10,19,20], which, based on our results, suggests the existence of an As-rich or As-O-rich interface. This could be the result of either the formation of As oxides, which can decompose, leaving behind elemental As, or a high density of As-O bonds at the interface. As for GaAs/$\beta$-$Ga_2O_3$ heterostructures grown epitaxially or assembled via semiconductor grafting, both type-I [9] and type-II [8,21] alignments have been reported. Overall, our findings offer insight for tailoring interface properties to specific applications by controlling the interfacial stoichiometry.

However, additional factors, such as defects and doping, could also affect the properties of the fabricated heterostructures. While our results indicate similar band alignments for As-rich and As-O-rich interfaces, their structural features differ. Specifically, As-dimers in the As-rich interface could be detrimental to the device performance, as they are associated with Fermi level pinning [37,54]. Although we did not observe mid-gap states in any of the studied models, Colleoni *et al*. [36] demonstrated that excess charge in systems containing such dimers can induce Fermi level pinning and mid-gap states through transitions between As-dimers and As dangling bonds. These effects must be considered in the design of GaAs/$Ga_2O_3$ heterostructures.

In summary, we investigated the influence of interfacial chemistry on the band alignment between $Ga_2O_3$ and GaAs using hybrid DFT calculations. Our results reveal that Ga-O-rich interfaces produce large VBOs and type-II alignment, whereas As-rich and As-O-rich interfaces yield smaller VBOs and type-I alignment. These findings apply to both amorphous and crystalline $\beta$-$Ga_2O_3$, providing guidance for the rational design of GaAs/$Ga_2O_3$ heterostructures with tailored electronic properties.

This research was supported by the Brittany Region under the "Stratégie d'Attractivité Durable (SAD)" funding and by the “France 2030” program of the French National Research Agency, NAUTILUS Project (Grant no. ANR-22-PEHY-0013). We acknowledge the EuroHPC Joint Undertaking for awarding this project access to the EuroHPC supercomputer LUMI, hosted by CSC (Finland) and the LUMI consortium through a EuroHPC Regular Access call. The work was granted access to the HPC resources of TGCC/CINES under the allocation A0180911434 made by GENCI.

The data that support the findings of this article are openly available [55].

*Contact author: sofia.apergi@insa-rennes.fr
†Contact author: laurent.pedesseau@insa-rennes.fr

*Contact author: sofia.apergi@insa-rennes.fr
†Contact author: laurent.pedesseau@insa-rennes.fr

*Contact author: sofia.apergi@insa-rennes.fr
†Contact author: laurent.pedesseau@insa-rennes.fr

*Contact author: sofia.apergi@insa-rennes.fr
†Contact author: laurent.pedesseau@insa-rennes.fr